\documentclass[aps,pra,showpacs,twocolumn,superscriptaddress,nolongbibliography]{revtex4-2}
\usepackage{graphicx}
\usepackage{amsmath}
\usepackage{dcolumn}
\usepackage{bm}
\usepackage[colorlinks=true, citecolor=blue,allcolors=blue]{hyperref}

\begin{document}
\title{Two-path inter\-ference in resonance-enhanced few-photon ionization of Li atoms} 

\author{ B.P.~Acharya}
\affiliation{Physics Department and LAMOR, Missouri University of Science \& Technology, Rolla, MO 65409, USA}

\author{S.~Dubey}
\affiliation{Physics Department and LAMOR, Missouri University of Science \& Technology, Rolla, MO 65409, USA}

\author{K.L.~Romans}
\affiliation{Physics Department and LAMOR, Missouri University of Science \& Technology, Rolla, MO 65409, USA}

\author{A.H.N.C.~De Silva}
\affiliation{Physics Department and LAMOR, Missouri University of Science \& Technology, Rolla, MO 65409, USA}

\author{ K.~Foster}
\affiliation{Physics Department and LAMOR, Missouri University of Science \& Technology, Rolla, MO 65409, USA}

\author{O.~Russ}
\affiliation{Physics Department and LAMOR, Missouri University of Science \& Technology, Rolla, MO 65409, USA}

\author{K.~Bartschat}
\affiliation{Department of Physics  and Astronomy, Drake University, Des Moines, Iowa 50311, USA}

\author{N.~Douguet}
\affiliation{Department of Physics, Kennesaw State University, Kennesaw, Georgia 30144, USA}

\author{D.~Fischer}
\affiliation{Physics Department and LAMOR, Missouri University of Science \& Technology, Rolla, MO 65409, USA}

\date{\today}

\begin{abstract}
We investigate the resonance-enhanced few-photon ionization of atomic lithium by 
linearly polarized light whose frequency is tuned near the $2s-2p$ transition. Considering the direction of 
light polarization orthogonal to the quantization axis, 
the process can be viewed as an atomic ``double-slit experiment'' where the 2$p$ states with magnetic quantum numbers $m_\ell=\pm1$ act
as the slits. In our experiment, we can virtually close one of the two slits by preparing lithium in one of 
the two circularly polarized $2p$ states before subjecting it to the ionizing radiation.
This allows us to extract the inter\-ference term between the two pathways and
obtain complex phase information on the final state. The experimental results show very good agreement with numerical solutions of the time-dependent Schr\"{o}dinger equation. The validity of the two-slit model is also analyzed theoretically using a time-dependent perturbative approach.
\end{abstract}

\maketitle

\section{Introduction}

Two-path inter\-ference is one of the most intriguing and intensely studied phenomena in physics. 
It was first demonstrated in 1801 for optical light by Thomas Young in his well-known double-slit 
experiment \cite{Young1802}. The historic importance of this experiment for the development of quantum theory is hard 
to overstate, because it reveals the wave nature of massive particles such as 
electrons \cite{Davisson1928,Joensson1961}, atoms \cite{Keith1991}, and even large 
molecules \cite{Arndt1999}, thereby supporting de Broglie's hypothesis of wave-particle 
duality \cite{deBroglie1924}. Until today, this phenomenon has not lost its appeal, 
and it has been observed in numerous systems. On one hand, it allows to extract 
phase information on wave functions, which is commonly not directly observable. 
On the other hand, it is exploited in many quantum-control schemes, because the 
manipulation of the relative amplitudes of the two pathways makes it possible to control 
the final state with high sensitivity. In atomic and molecular scattering processes, 
examples include well-known effects like Feshbach, shape,  and Fano 
resonances (e.g.\ \cite{Feshbach1958,Fano1961,Schulz1973,Chin2010}), 
or atomic-scale double-slits formed by diatomic molecules exhibiting inter\-ferences 
in differential ionization cross sections due to ion \cite{Stolterfoht2001,Misra2004,Egodapitiya2011,Zhang2014}, 
electron \cite{MilneBrownlie2006,Li2018}, or photon impact \cite{Cohen1966,Kunitski2019}.

Multiphoton ionization processes of single atoms expose two- and multi\-path inter\-ferences 
in a particularly clean way, because of the well-defined energy and limited angular-momentum 
transfer in photon absorption reactions. A prominent example  is RABBITT 
(Reconstruction of Attosecond Beating by Inter\-ference of Two-photon Transitions) spectroscopy  
\cite{Muller2002,Kluender2011,Isinger2017,Bharti2021}, which has become the standard tool to 
characterize extreme-ultraviolet (XUV) attosecond pulse trains and allows the study 
of atto\-second atomic dynamics in the time domain. Two-color ionization schemes using (lower) 
harmonic radiation \cite{Yin1992,Ehlotzky2001,Brif2010,Giannessi2018} enable the coherent 
control of the reactions' final state via two-path inter\-ferences. Recently, other schemes 
have been considered, where double-slit structures in so-called Kramers-Henneberger states 
emerge through the distortion of a bound state by an external field, once again resulting in inter\-ference 
patterns \cite{He2020}. Two-path inter\-ference has not only been observed in laser pulses 
but also using two mutually incoherent (i.e., without relative phase lock) continuous-wave (cw) 
lasers in the two-photon ionization of \hbox{rubidium} atoms \cite{Pursehouse2019}, where the 
photon energies are tuned to two different resonances.

\begin{figure}
\centering
\includegraphics[width=1\linewidth]{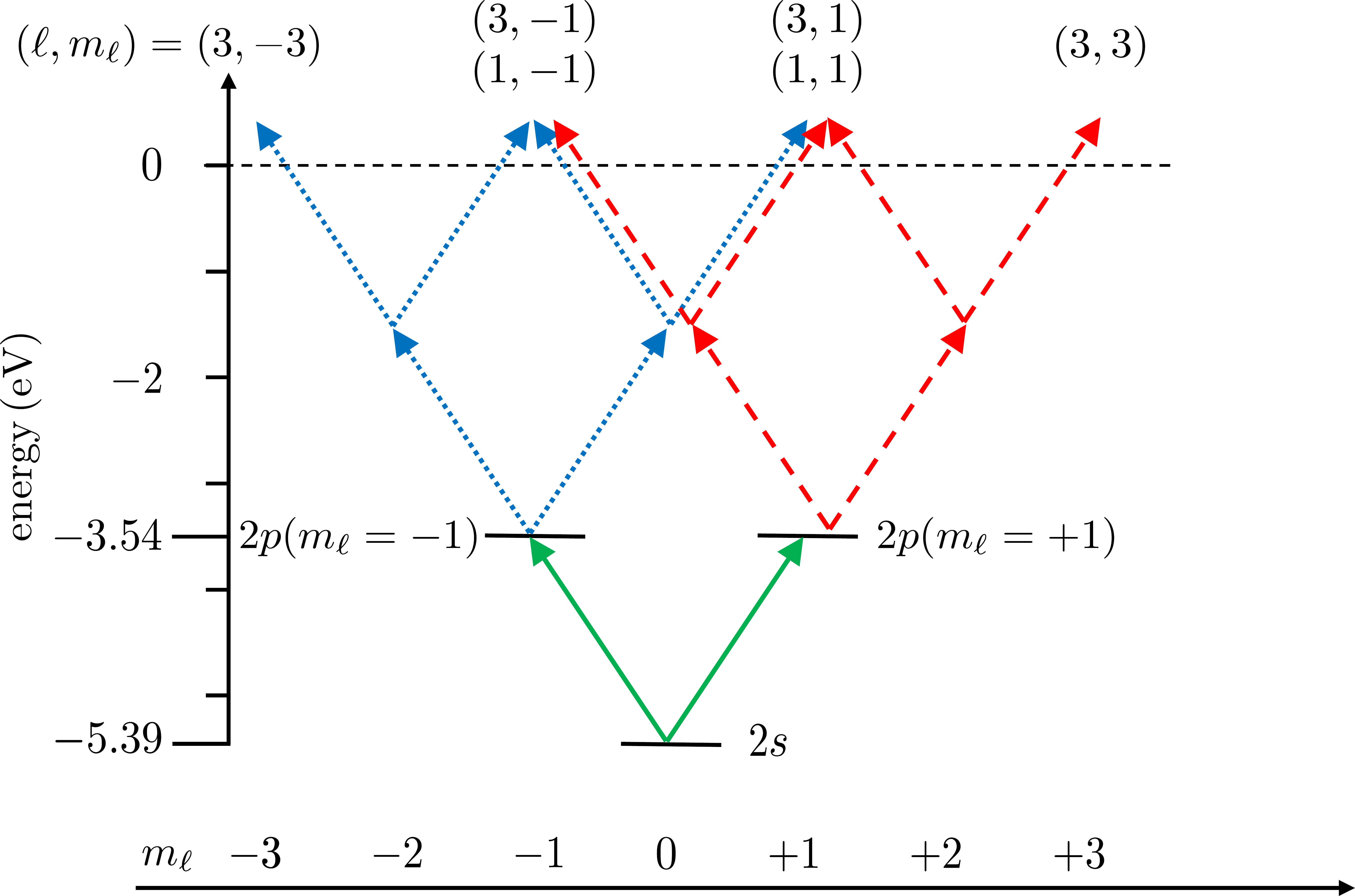}
\caption{Few-photon ionization scheme in lowest-order perturbation theory. 
The ionization pathways from the 2$p$ state with $m_\ell=+1$ and $-1$ are shown as 
red dashed and blue dotted arrows, respectively. The 2$s$ ionization corresponds 
to the superposition of both.\label{fig:scheme}}
\end{figure}

In the present study, two-path inter\-ference occurs in the ground-state ionization of 
lithium exposed to single-color femtosecond laser pulses, which are linearly polarized 
in the $y$~direction. The laser spectrum has its center wavelength at 660\,nm and partially overlaps 
with the $2s-2p$ resonance at 671\,nm. For the quantization axis chosen as the $z$~direction, 
the absorption of a single photon results in the excitation to the 2$p$ state coherently 
populating the two magnetic sub\-levels with $m_\ell=+1$ and $-1$, respectively. 
These two eigenstates resemble 
the two ``slits'' in analogy to Young's double-slit scheme (see Fig.~\ref{fig:scheme}). 
From these two excited levels, the atom is ionized without further resonance enhancement 
by the absorption of two more photons from the same laser pulse.  The final result is a 
super\-position of electronic $p$ and $f$ continuum waves.

It is important to note that the distinction of these two pathways relies on the choice 
of the quantization direction. However, this choice is motivated by the experimental 
capability of preparing the atoms selectively in one of the two excited and polarized 
magnetic sub\-levels of the 2$p$ state before exposing them to the femtosecond laser pulse. 
This enables us to measure not only the final intensity of the two inter\-fering pathways, 
which corresponds to the differential cross sections for the ionization of the 2$s$ state, 
but also the intensity of each pathway individually via 2$p$ 
ionization. This fact can approximately be expressed as (see appendix \ref{ap:two-path-der} for a derivation
and discussion of the validity criteria)
\begin{equation}
\begin{aligned} 
\left| \mathcal{A}_{2s}(\bm p)\right|^2 \approx\alpha\left| \mathcal{A}_{2p}^{+}(\bm p) + \mathcal{A}_{2p}^{-}(\bm p)\right|^2,\\
\end{aligned}
\label{eq:interference}
\end{equation}
where $\mathcal{A}_{2s}(\bm p) $, $\mathcal{A}_{2p}^{+}(\bm p) $, and $\mathcal{A}_{2p}^{-}(\bm p)$ 
represent the ionization amplitudes for a photoelectron with asymptotic momentum 
$\bm p$ from the initial 2$s\,(m_\ell\!=\!0)$, 2$p\,(m_\ell\!=\!+1)$, and 2$p\,(m_\ell\!=\!-1)$ states, respectively, while $\alpha$ is a
real factor. Equation \eqref{eq:interference} holds in our case under the assumptions that (i) 
the pulse is sufficiently long and the light frequency is tuned near the $2s-2p$ transition, such that only 
resonant transitions to the $2p$ states through the first photon absorption contribute (virtual excitations can be neglected), 
and (ii) the field is sufficiently weak that the ionization amplitude can be expressed through a perturbative expansion (even beyond the lowest order). 
While the transfer between $2s$ and $2p$ was adiabatic in our case, we found Eq.~\eqref{eq:interference} 
to remain fairly accurate even for non\-adiabatic transfer (see appendix \ref{ap:two-path-der} for more details). 
In addition, the experimental observations and 
calculations will be shown to be in excellent agreement, thereby providing further evidence of the validity of Eq.~\eqref{eq:interference}. 
Using this approach, we obtain a direct and intuitive way to extract 
the complex phase difference between $\mathcal{A}_{2p}^{+}$ and $\mathcal{A}_{2p}^{-}$ 
as a function of the electron emission angle, thereby revealing the effect of 
the orientation of the initial electron orbital angular momentum on the final state's phase.

\section{Methods}

The experimental technique and the theoretical method are identical to those reported in 
previous studies on very similar systems \cite{Silva2021,Silva2021b,Acharya2021}. 
Therefore, only some key features are repeated here and parameters specific to the present study are mentioned. 

Lithium atoms are cooled and confined in a volume of about 1\,mm diameter in a 
near-resonant all-optical atom trap (AOT) \cite{Sharma2018} with a fraction of 
about 25\,\% being in the polarized excited $2p\,(m_\ell\!=\!+1)$ state and 
about 75\,\% in the 2$s$ ground state. The atoms are ionized in the field 
of a femtosecond laser based on a Ti:Sa oscillator with two non\-collinear 
optical parametric amplifier (NOPA) stages. For the present study, 
the laser wavelength was chosen to center at 660\,nm with pulse durations (FWHM of intensity)
of about 65\,fs and a peak intensity of about 3$\times$10$^{10}$\,W/cm$^2$. 
The three-dimensional electron momentum vectors are measured with a resolution 
of about 0.01\,a.u.\ \cite{Thini2020} in a reaction microscope, which is described in detail in \cite{Hubele2015,Fischer2019}. 
It is important to note that this experimental setup enables us to obtain differential 
cross-normalized data for the ionization of the 2$s$ and the 2$p$ initial states simultaneously.

In our theoretical model, the lithium atoms are approximated as a single active 
electron moving in the field of a 1$s^2$ ionic core.  The latter is described by 
a static Hartree potential \cite{Albright1993,Schuricke2011}, which is 
supplemented by phenomenological terms to account for the core polarizability 
as well as exchange between the valence electron and those in the core~\cite{Silva2021}. The (complex) final-state wave 
function is obtained after propagating the initial state in time by 
numerically solving the time-dependent Schr{\"o}dinger equation (TDSE).

\section{Results and Discussion}

\begin{figure}
\centering
\includegraphics[width=0.8\linewidth]{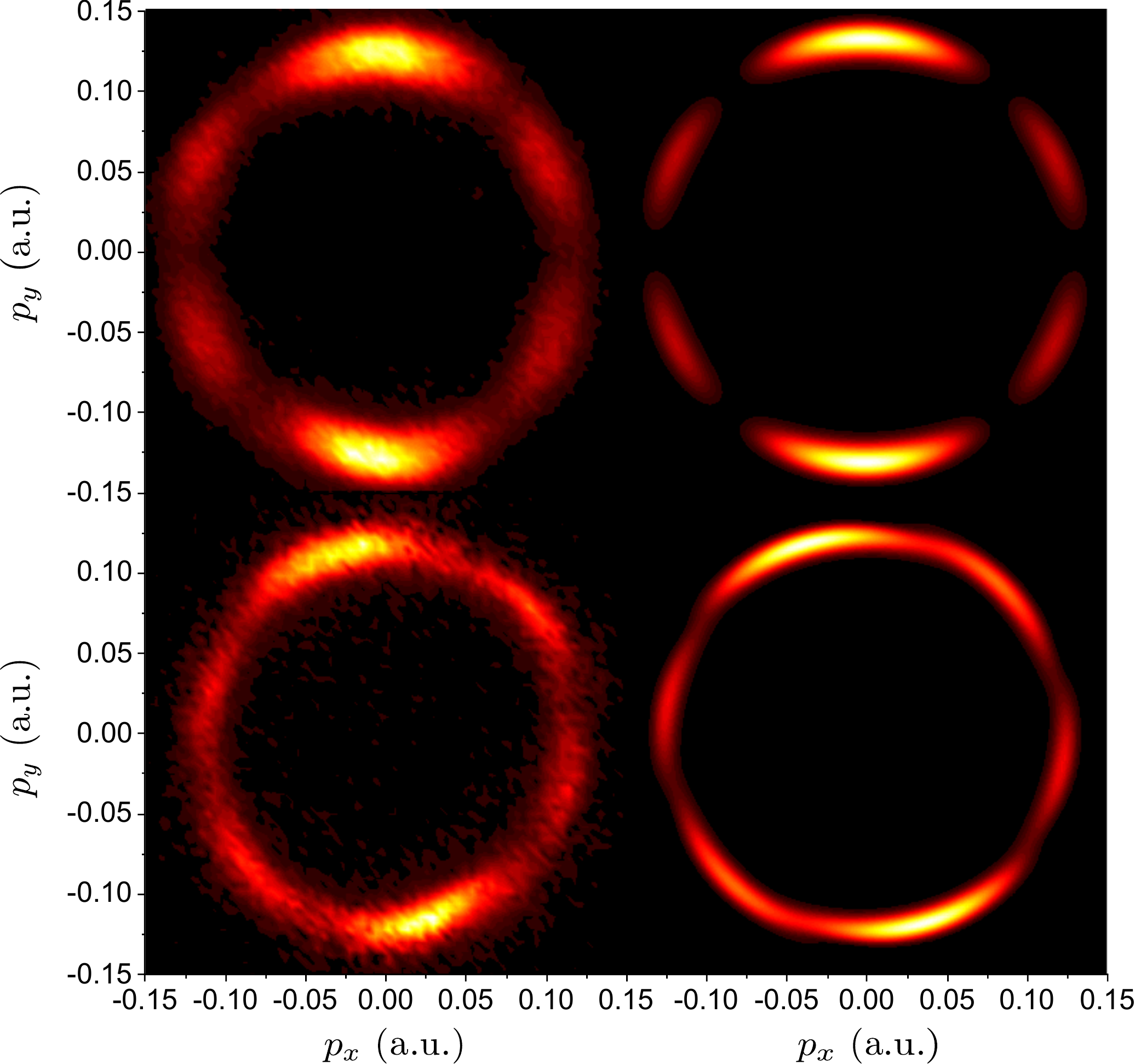}
\caption{Experimental (left) and theoretical (right) photo\-electron momentum distributions 
projected onto the $xy$ plane for few-photon ionization of the 2$s$ (top row) and 2$p\,(m_\ell\!=\!+1)$ 
(bottom row) initial states by linearly polarized laser pulses of 65\,fs duration with a center 
wavelength of 660\,nm and a peak intensity of 3.1$\times10^{10}$\,W/cm$^2$. 
The laser polarization direction is along the $y$ axis (i.e., vertical), while the atomic 
initial orbital angular momentum is oriented in the $z$ direction (i.e., perpendicular to the 
drawing plane). \label{fig:expspectra}}
\end{figure}

In the present study, lithium atoms in the 2$s$ ground state and 2$p$ excited state 
are ionized in a laser field with a central wavelength of 660\,nm at intensities well below 10$^{11}$\,W/cm$^2$.  
This situation corresponds to Keldysh parameters $\gamma>20$, and hence the system is expected 
to be well described in a multi\-photon picture. The two initial states are ionized  by the 
absorption of (at least) three (from $2s$) or two (from $2p$) photons, respectively, resulting in a final electron 
energy of about 200\,meV. The measured and calculated electron momentum spectra shown 
in Fig.~\ref{fig:expspectra} are in excellent agreement with each other. 
Before proceeding to the analysis of the two-path inter\-ference introduced above, 
two important features of the data should be mentioned, even though they were already reported
previously in several recent studies~\cite{Silva2021,Acharya2021}. 

First, while the photo\-electron momentum distributions (PMDs) for 2$s$ ionization exhibit 
reflection symmetry with respect to the laser electric field direction 
(the vertical direction in the momentum spectra shown in Fig.~\ref{fig:expspectra}), 
this symmetry is broken for ionization of the polarized 2$p$ state. Consequently, the main 
electron emission direction appears to be shifted. The dependence of this phenomenon, known as 
magnetic dichroism, on the laser wavelength and intensity was 
recently investigated by Acharya {\it et al.}~\cite{Acharya2021}. In this earlier study, 
these asymmetries were explained in a partial-wave picture.  They were traced back 
to a non\-vanishing mean orientation of the final electron orbital angular 
momentum $\left< m_\ell\right>\neq 0$.  This \hbox{``remnant''} of the initial 
target polarization is partially preserved throughout the ionization process. 
The angular shifts and observed asymmetries are a result of the inter\-ference 
between (phase-shifted) partial waves with different~$m_\ell$. 

Second, the azimuthal photo\-electron angular distributions (PADs) for 2$p$ ionization 
feature six peaks.  As discussed below, this indicates beyond lowest-order contributions to 
the ionization cross section. Generally, the dependence of the differential 
cross section on the \hbox{azimuthal} angle $\varphi$ is given by~\cite{Acharya2021}
\begin{equation}
 \frac{\mathrm{d}\sigma}{\mathrm{d}\Omega}=\left| \sum_{m_\ell} c_{m_\ell}e^{im_\ell\varphi}\right|^2,
 \label{eq:dcs}
\end{equation}
where the factors $c_{m_\ell}$ relate to the complex amplitudes of the partial waves.  
In lowest-order perturbation theory (LOPT), only the shortest pathways to the final state (i.e., the absorption of only the minimum 
number of photons needed to reach the final photoelectron energy) are considered. For the present initial 2$p\,(m_\ell\!=\!+1)$ state, 
this corresponds to two-photon absorption.  In the electric dipole 
approximation, this results in partial waves with $m_\ell=-1$, $+1$, and $+3$ contributing 
to the final state (cf.~Fig.~\ref{fig:scheme}). 
For this set of dipole-allowed $m_\ell$ values, therefore, the above expression results in a 
photo\-electron angular distribution with no more than four peaks, in contrast to the six peaks 
observed in both the experiment and the \emph{ab initio} calculation. This evident violation of LOPT 
close to the \hbox{$2s - 2p$} resonance was reported and discussed in our previous study as well: 
It is explained by the coupling between the 2$s$ and 2$p$ states in the external field 
giving rise to adiabatic population transfer between these two states and resulting in a contribution 
of $m_\ell=-3$ to the final state. Accounting for this additional pathway, the expression 
in Eq.~(\ref{eq:dcs}) allows for angular distributions with up to six peaks, 
which is consistent with experiment and calculation.

\begin{figure}
\centering
\includegraphics[width=1\linewidth]{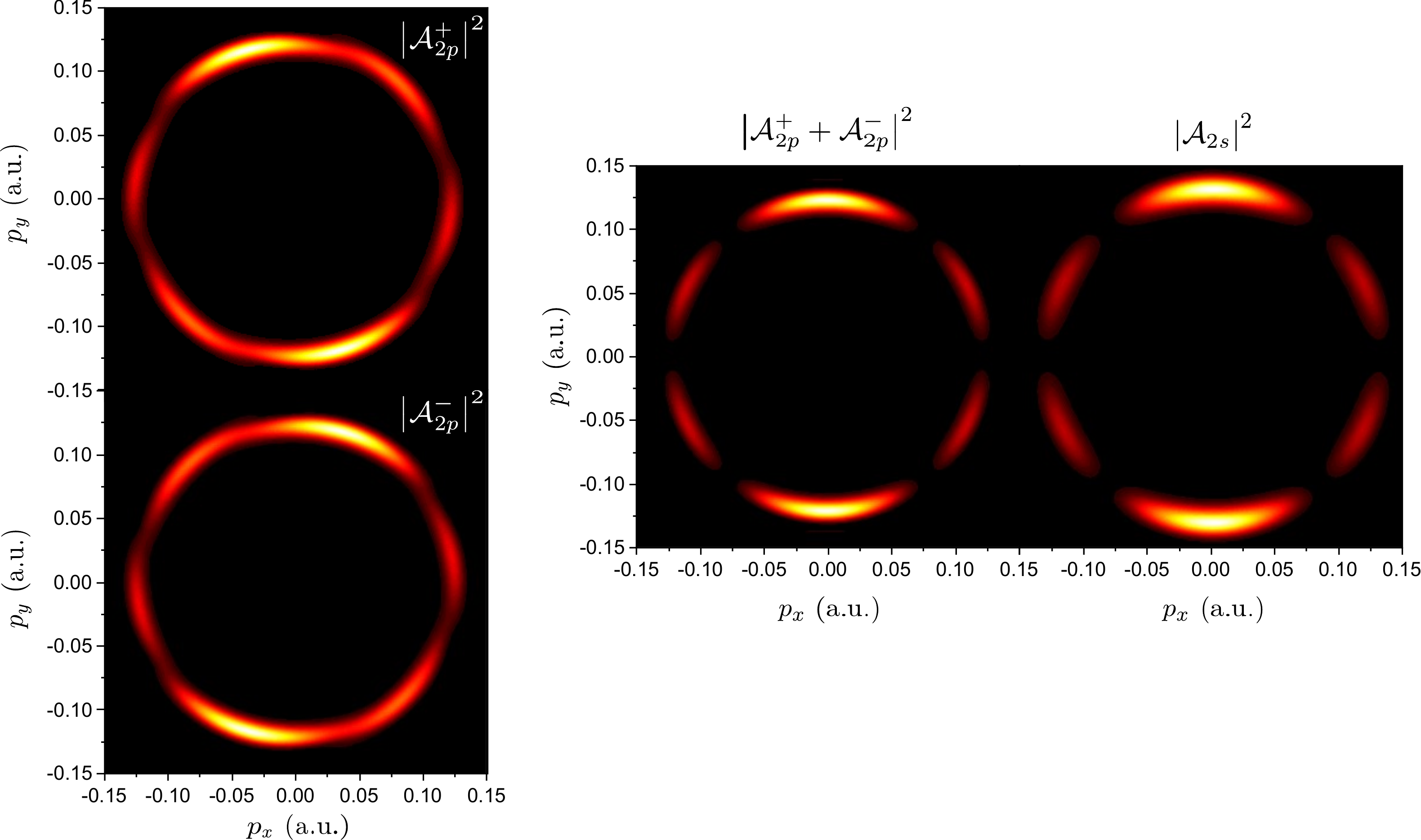}
\caption{Absolute square of the calculated wave functions $\mathcal{A}_{2p}^+$  and $\mathcal{A}_{2p}^-$ (left), 
of their coherent sum $\mathcal{A}_{2p}^++\mathcal{A}_{2p}^-$ (center), and of $\mathcal{A}_{2s}$ (right) 
in the $xy$ plane in momentum space.  See text for details.  \label{fig:th2p}}
\end{figure}

The validity of the two-path inter\-ference expression given in Eq.~(\ref{eq:interference}) 
can be tested by using our theoretical description. In a first step, the $\mathcal{A}_{2p}^{-}$ and $\mathcal{A}_{2p}^{+}$ amplitudes for the 
ionization of the 2$p$ initial states with $m_\ell=-1$ and $+1$, respectively, 
are calculated. Their absolute squares, corresponding to the differential ionization 
cross sections, are shown as a function of the photo\-electron momentum in 
Fig.~\ref{fig:th2p} (left) considering only electron emission in the $xy$ plane
(i.e., for a polar angle $\vartheta=90^\circ$). Due to symmetry considerations, 
the systems with opposite initial orbital angular momentum $m_\ell\!=\!+1$ 
and $-1$ are mirror images of one another with the mirror plane spanned 
by the laser polarization direction (the $y$ axis) and the direction of 
the initial atomic polarization (the $z$ axis). Specifically, $\mathcal{A}_{2p}^+(\sigma_{yz}\bm p)=\mathcal{A}_{2p}^-(\bm p)$, 
where $\sigma_{yz}$ is the reflection operator through the $yz$ plane.

In the second step, Eq.~(\ref{eq:interference}) is tested by comparing the calculated momentum 
distribution for 2$s$ ionization  (corresponding to $\left|\mathcal{A}_{2s}\right|^2$) with the 
intensity of the super\-position of the two amplitudes for 2$p$ ionization $\left|\mathcal{A}_{2p}^++\mathcal{A}_{2p}^-\right|^2$. 
The PMDs obtained by both methods are shown in Fig.~\ref{fig:th2p} 
(right and center, respectively).  They are in overall very good agreement indeed, although 
the $\left|\mathcal{A}_{2s}\right|^2$ distribution has a slightly larger diameter. This small 
discrepancy is a result of slightly different (by approximately 10\,\%) photo\-electron energies 
for 2$s$ and 2$p$ ionization, because the wavelength of the ionizing field is off 
the $2s - 2p$ resonance by about 10\,nm.

The discussion above shows that the final momentum distribution for 2$s$ 
ionization (which corresponds to $\left|\mathcal{A}_{2s}\right|^2$) can be calculated, to a good 
approximation, from the  $\mathcal{A}_{2p}^+$  ionization amplitude by exploiting 
Eq.~(\ref{eq:interference}) and the mirror symmetry between $\mathcal{A}_{2p}^+$ and $\mathcal{A}_{2p}^-$. 
Evidently, this is not possible with the experimental data, because only the 
absolute square of the final-state wave function $|\mathcal{A}_{2p}^+|^2$ is directly measured. 
However, the relative phase between $\mathcal{A}_{2p}^-$ and $\mathcal{A}_{2p}^+$ 
can be extracted from Eq.~(\ref{eq:interference}) by 
reversing the above procedure and solving 
for the phase difference. This yields
\begin{eqnarray}
\cos\Delta\phi(\bm p)=\frac{\left|\mathcal{A}_{2s}(\bm p)\right|^2-\alpha\left| \mathcal{A}_{2p}^{+}(\bm p)\right|^2-\alpha\left|\mathcal{A}_{2p}^{-}(\bm p)\right|^2}{2\alpha\left| \mathcal{A}_{2p}^{+}(\bm p)\right|\left| \mathcal{A}_{2p}^{-}(\bm p)\right|}, 
\label{eq:deltaphi}
\end{eqnarray}
where $\Delta\phi(\bm p)=\arg[\mathcal{A}_{2p}^{+*}(\bm p)\mathcal{A}_{2p}^{-}(\bm p)]$. The reconstruction of the 
phase difference $\Delta\phi$
in three-dimensional momentum space from experimental data using Eq.~(\ref{eq:deltaphi}) 
requires matching photo\-electron energies for 2$s$ and 2$p$ ionization, 
which is not strictly fulfilled for the present laser wavelength. 
The question arises whether, in spite of the small photo\-electron 
energy mismatch, Eq.~(\ref{eq:deltaphi}) is still applicable if 
only the dependence on the electron emission angle is considered 
(i.e., for the electron energy fixed at the peak energy). 
This can be the case, if the angular distributions do not significantly 
vary with small shifts of the photo\-electron energy. This is tested by comparing the 
calculated angular distributions for 2$s$ ionization (extracted from $\left|\mathcal{A}_{2s}\right|^2$) 
with the distribution obtained for the inter\-fering wave functions $\left|\mathcal{A}_{2p}^++\mathcal{A}_{2p}^-\right|^2$. 

\begin{figure}
\centering
\includegraphics[width=.9\linewidth]{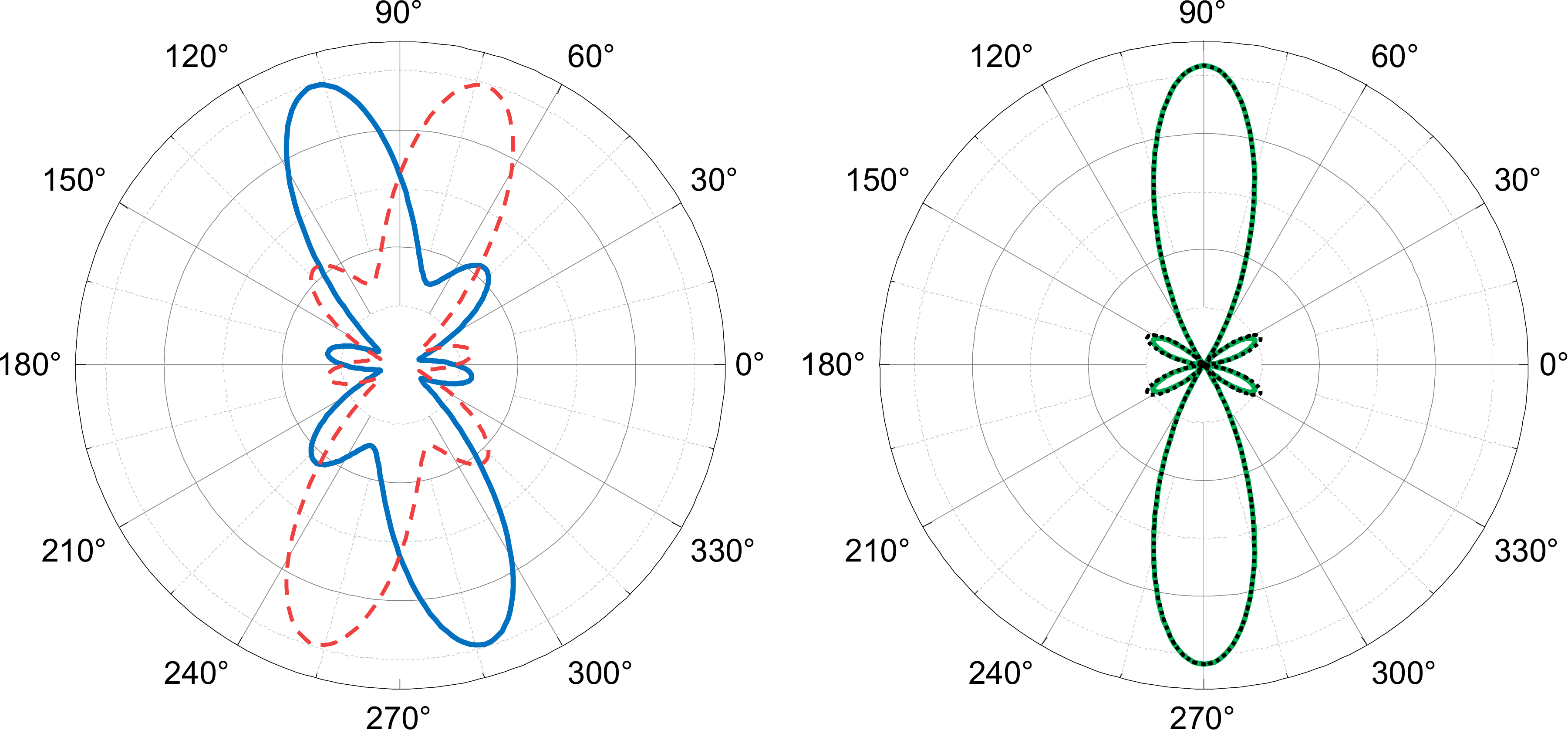}
\caption{Angular distributions extracted from the wave functions in Fig.~\ref{fig:th2p} 
as a function of the photelectron azimuthal angle $\varphi$. Left: angular distribution 
of $\left| \mathcal{A}_{2p}^{+}\right|^2$ (solid blue line) and $\left| \mathcal{A}_{2p}^{-}\right|^2$ (dashed red line). 
Right: $\left| \mathcal{A}_{2p}^{+}+\mathcal{A}_{2p}^-\right|^2$ (dashed black) and  $\left| \mathcal{A}_{2s}\right|^2$ (solid green). \label{fig:thangdist}}
\end{figure}

The corresponding angular distributions are shown in Fig.~\ref{fig:thangdist}. Indeed, 
the angular spectrum obtained from the inter\-fering wave functions closely resembles 
the distribution calculated for 2$s$ ionization. 
There is only a small deviation in the relative intensity of the main peak in the polarization 
direction and the side peaks. Therefore, we conclude that Eq.~(\ref{eq:deltaphi}) makes it possible 
to extract the phase difference $\Delta\phi$ (to a good 
approximation) as a function of the azimuthal angle for the peak photo\-electron energy.

Before Eq.~(\ref{eq:deltaphi}) can be employed to calculate the phase difference $\Delta\phi$ 
from the experimental data, the factor~$\alpha$ must be determined. Here we can borrow 
an idea from Young's double-slit experiment, where we know that the total flux is conserved, i.e., 
the total intensity equals the sum of the intensities going through each slit 
individually. In our case, this means that the momentum-integrated interference term in Eq. \eqref{eq:interference} should vanish, i.e., 
\begin{equation}
\int\mathrm{d}^3p\left|\mathcal{A}_{2p}^{+}(\bm p)\right|\left|\mathcal{A}_{2p}^{-}(\bm p)\right|\cos[\Delta\phi(\bm p)]=0.
\label{eq:inter_averaged}
\end{equation}
Equation \eqref{eq:inter_averaged} holds since the azimuthal dependence of the interference is expressed as the superposition
of terms of the form $|c_n||c_m|\cos{[(n-m)\phi}]$, where $n\ne m$, and hence $\int_0^{2\pi}d\phi\cos{[(n-m)\phi]}=0$. 
Therefore, the inter\-ference term does not contribute to the total intensity,  
and the factor~$\alpha$ is readily found as
\begin{equation}
\alpha=\frac{\int\mathrm{d}^3p\left|\mathcal{A}_{2p}^+(\bm p)\right|^2 +\int\mathrm{d}^3p\left|\mathcal{A}_{2p}^-(\bm p)\right|^2}{\int\mathrm{d}^3p \left| \mathcal{A}_{2s}(\bm p)\right|^2}.
\label{eq:normlization}
\end{equation}

\begin{figure}
\centering
\includegraphics[width=1\linewidth]{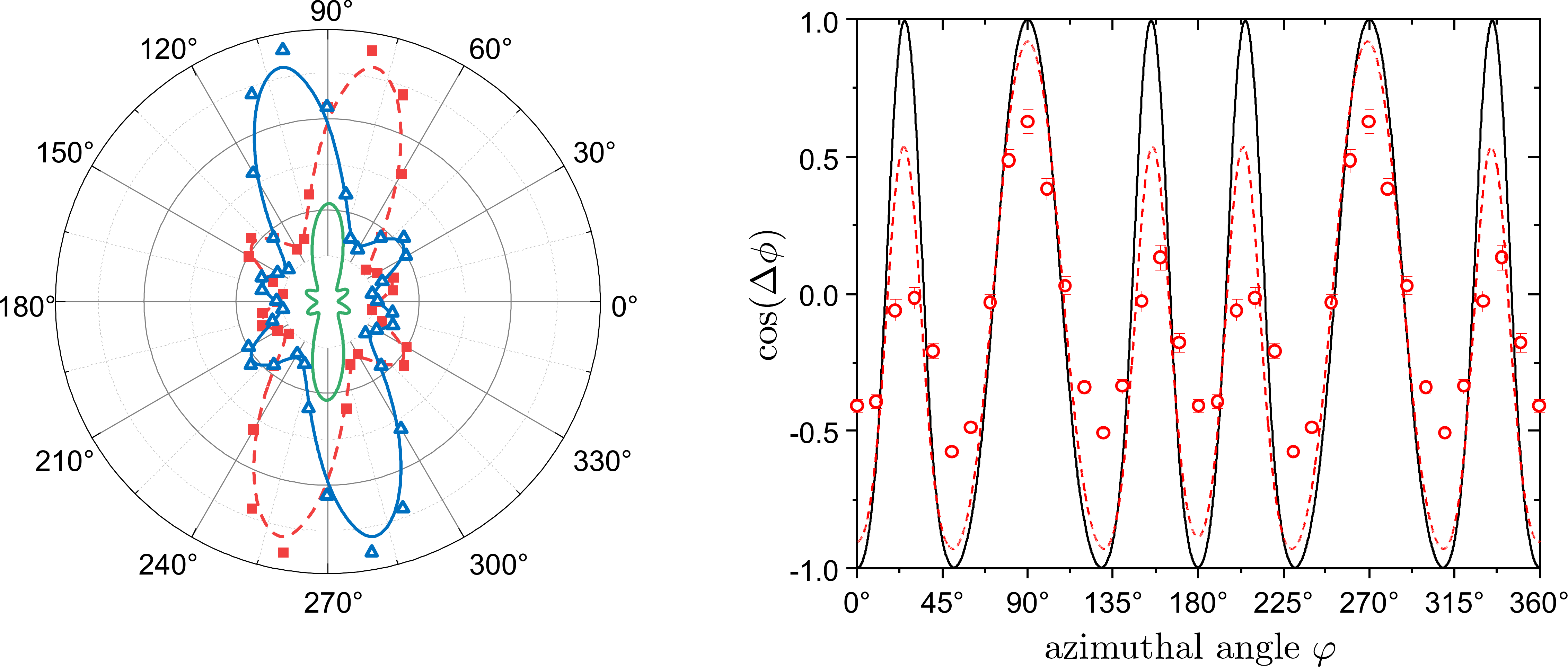}
\caption{Left: Experimental PADs as a 
function of the azimuthal angle $\varphi$ for the few-photon ionization 
of lithium initially in the 2$s$ (solid green line) and 2$p$ state with $m_\ell\!=\!+1$ 
(blue solid line and open triangles) and $-1$ (red dashed line and solid squares). 
The lines are inter\-polating splines to guide the eye. Right: Experimental (red open circles) and theoretical (solid line) 
cosine of the phase difference as a function of the photo\-electron 
azimuthal angle~$\varphi$. The dashed red line is derived from the theoretical angular distributions 
that were convolved with the experimental resolution (see text). \label{fig:expphase}}
\end{figure} 

The experimental PADs are shown in Fig.~\ref{fig:expphase} (left). 
While the distributions for the ionization for the 2$s$ and the 2$p\,(m_\ell\!=\!+1)$ 
initial states are measured directly in our experiment, the data for 2$p\,(m_\ell\!=\!-1)$ 
are obtained by reflecting the data for the opposite target polarization on the laser 
polarization axis. Using these angular distributions, the cosine of the phase difference is 
calculated with Eq.~(\ref{eq:deltaphi}), plotted in Fig.~\ref{fig:expphase} (right), and 
compared to the theoretical predictions. 

The distribution features six crests and troughs 
whose positions agree very well between theory and experiment. However, some discrepancies 
in the magnitude persist. While the calculated curve reaches the maximum and minimum 
values of +1 and $-1$, respectively, the oscillation is weaker in the experimental data. 
Generally, a value of $+1$ for $\cos\Delta\phi$ corresponds to maximum constructive 
inter\-ference, which is expected at angles where the angular distribution for 2$s$ 
ionization has a local maximum. Correspondingly, $\cos\Delta\phi=-1$ means complete 
destructive inter\-ference, which should occur at local minima in the differential 2$s$ 
ionization data. 

There are two effects that might blur these inter\-ferences in the 
experimental data: (1)~There is a small, but non\-negligible experimental angular 
uncertainty. The influence of this effect is shown by the red dashed line in the figure, 
which represents the $\cos\Delta\phi$ distribution derived from the theoretical 
angular distributions convolved with the experimental angular resolution (ca.~12$^\circ$ FWHM). 
(2)~The experimental data represent an average over a laser 
intensity range, as already discussed above. As the angular distributions are not 
entirely independent of the laser intensity (see Fig.~\ref{fig:expspectra}), this will also 
result in a blurring of the data. 

\begin{figure}
\centering
\includegraphics[width=0.8\linewidth]{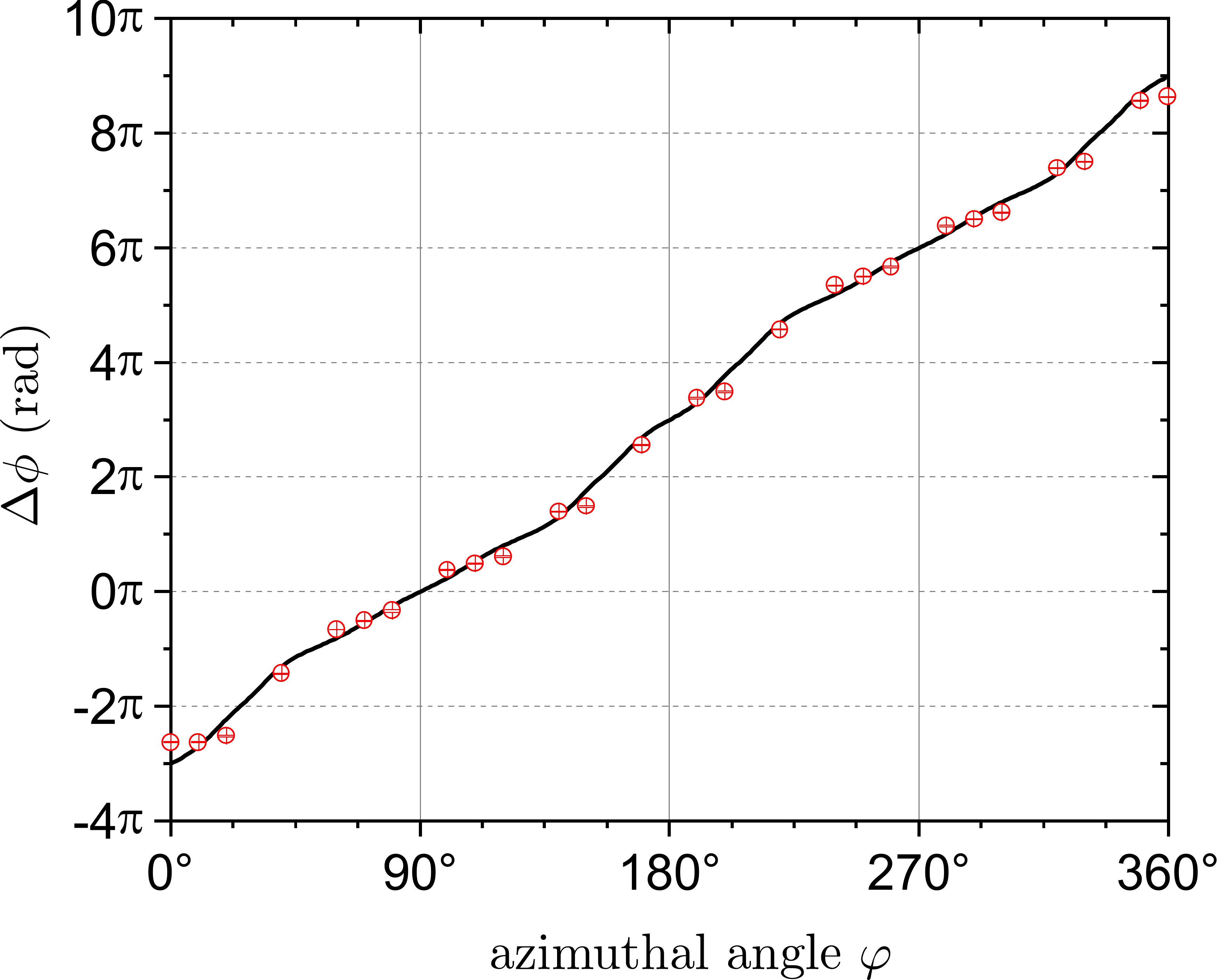}
\caption{Theoretical (line) and experimental (open circles) phase difference of the final state wave functions after two-photon 
ionization of the 2$p$ initial state with $m_\ell=+1$ and $-1$ as a function of 
the photoelectron azimuthal angle.  \label{fig:psiplusphase}}
\end{figure} 

The $\varphi$-dependence of the phase difference $\Delta\phi$ of the final-state 
wave functions can be derived from the data shown in Fig.~\ref{fig:expphase} (right) 
and is presented in Fig.~\ref{fig:psiplusphase}. It should be noted that extracting this 
phase difference is somewhat ambiguous due to the oscillatory behavior of the cosine function. 
In the present case we made the additional assumption that the phase increases monotonically 
with the azimuthal angle $\varphi$. The phase difference is $\Delta\phi=0$ at 
an angle of 90$^\circ$ (i.e., in the $y$ direction) due to the symmetry $\Delta\phi(90+\varphi)=-\Delta\phi(90-\varphi)$. 
Overall, the phases obtained from the 
experimental data are consistent  and in very good agreement with the theoretical prediction, 
thereby supporting the validity of the two-path inter\-ference picture developed here. The
remaining deviations are attributed to the blurring effects discussed in the previous paragraph. 
It is important to note that the method used above does not allow to unambiguously 
extract the individual phases of the $\mathcal{A}_{2p}^+$ and $\mathcal{A}_{2p}^-$ amplitudes without using 
further assumptions, for example, when the angular dependence of the wave functions is 
expressed as a super\-position of a limited set of spherical harmonics, as was done in Eq.~(\ref{eq:dcs}).
 
\section{Conclusions and Outlook}

We studied the details of electron emission in few-photon ionization of lithium 
atoms initially either in the 2$s$ ground state or in the polarized 2$p\,(m_\ell\!=\!+1)$ excited 
state by radiation close to the $2s-2p$ resonance. We exploited the fact that the $2s$ state can 
be ionized through two possible pathways, specifically via the 2$p$ resonance with 
either $m_\ell\!=\!+1$ or $-1$. These two pathways inter\-fere in the final state and resemble 
a double-slit. Because our experiment allows us to obtain the differential cross sections 
for the 2$s$ and the 2$p$ initial states separately, we are able to measure the final wave 
with both ``slits'' open, or with one ``slit'' closed. Therefore, the data make it possible to extract
the inter\-ference term, thereby providing information on the relative phase of the two pathways. 
The experimentally obtained phase differences are in good agreement with our theoretical predictions.

Moreover, several interesting features are observed in the present data, 
which were reported for similar systems in preceding studies: First, the photo\-electron 
angular distributions after ionization of the polarized 2$p$ state are not symmetric 
with respect to the laser polarization.  Instead, the peaks are shifted. The wavelength and 
intensity dependence of this effect, known as magnetic dichroism, was systematically 
studied in~\cite{Acharya2021}. Second, the peak structures in the present 
angle-differential spectra are in direct contradiction to the predictions of lowest-order perturbation theory,
which is not applicable in resonant condition.

It is worth noting that the present method is not the only way to access information 
regarding the final state's phase. Among the many possible approaches, a straight\-forward one is 
to fit the angular distributions with model functions 
described by a superposition of partial waves as expressed in Eq.~(\ref{eq:dcs}). 
In some cases, this makes it possible to extract the relative phases 
between the complex amplitudes of partial waves contributing to the final state. 
For single-photon ionization, such \emph{complete} studies were 
pioneered in the 1990s using polarized atomic targets \hbox{\cite{Pahler1992,Becker1998,Godehusen1998}}. 
In the multiphoton ionization regime, phase information was obtained by ionizing atoms with 
elliptically polarized light \cite{Dulieu1995,Wang2000} in a very similar way. In contrast, the present scheme, 
which exploits the resonance enhancement through two magnetic sub\-levels, provides direct, 
complete, and intuitive access to the inter\-ference term and the final-state phase.

Two- or multi\-path inter\-ferences in few-photon ionization are well suited for quantum control schemes, 
if the relative phases and intensities of the different paths can be regulated (e.g., \cite{Giannessi2018}). 
It is interesting to conceive such a scheme for the present system. In fact, controlling the 
relative (complex) amplitudes of the transient 2$p\,(m_\ell\!=\!-1)$ and  2$p\,(m_\ell\!=\!+1)$ populations 
is experimentally straight\-forward. The transitions from the 2$s$ ground state to the two polarized 
excited 2$p$ levels are driven by left- and right-handed circularly polarized laser radiation, 
respectively, propagating in the $z$~direction. The superposition of these two fields with equal 
intensity and fixed relative phase corresponds to the linearly polarized light used in the present experiment. 
Changing the relative phase corresponds to a rotation of the polarization direction in 
the $xy$ plane. 

Furthermore, a change in the relative intensities can be achieved by introducing an ellipticity 
to the radiation. In the present scheme, quasi-monochromatic light is used and changes of the 
laser polarization would also affect the ionization steps after populating the resonant 2$p$ levels. 
However, the effect on the excitation process and the ultimate ionization could be de\-coupled by 
using bichromatic laser fields with a weak contribution close to the $2s - 2p$ resonance and a 
stronger contribution off resonance. Such an experiment would allow to prepare an atomic target 
in a coherent superposition of excited magnetic sub\-levels before ionizing it, thereby providing 
numerous possibilities to analyze and control the final state. 

\section*{Acknowledgments}
The experimental material presented here is based upon work supported by the National Science Foundation
under Grant \hbox{No.~PHY-1554776}.
The theoretical part of this work was funded by the NSF under
grants \hbox{No.~PHY-2012078} (N.D.), \hbox{PHY-1803844} and \hbox{PHY-2110023} (K.B.), and by the XSEDE 
supercomputer allocation \hbox{No.~PHY-090031}.

\appendix

\section{Derivation and conditions of applicability of the two-pathway formula}
\label{ap:two-path-der}

We now derive Eq.~\eqref{eq:interference} and discuss its domain of applicability.  
We consider the three-photon ionization amplitude $\mathcal{A}^{(3)}_{2s}(\bm p)$ for 
the system initially in the Li $|2s\rangle$ ground state under\-going a transition to 
a photoelectron with asymptotic momentum $\bm p$.
In the interaction picture, the amplitude is given by
\begin{eqnarray}
\mathcal{A}^{(3)}_{2s}(\bm p)&=&\left(-\frac{i}{\hbar}\right)^3\sum_{i,j}\int_{t_0}^{\infty}dt_1\int_{t_0}^{t_{1}}dt_{2}\int_{t_0}^{t_{2}}dt_{3} \nonumber\\
&\times& e^{i\frac{E}{\hbar}t_1}\hat{V}_{\bm k,j}(t_1) e^{-i\frac{E_{j}}{\hbar}(t_1-t_2)}\hat{V}_{j,i}(t_2)\nonumber\\\
&\times& e^{-i\frac{E_{i}}{\hbar}(t_2-t_3)}\hat{V}_{i,2s}(t_3)e^{-i\frac{E_{2s}}{\hbar}(t_3-t_0)}.
\label{eq:3phot}
\end{eqnarray}
In \eqref{eq:3phot}, $E_i$ are intermediate Li energies, $E_{2s}$ is the ground-state energy, 
$E={\hbar^2\bm k^2}/2m_e$ is the photoelectron energy, and $m_e$ is the electron mass. The summation 
runs over all bound and continuum electronic states.
The elements $\hat{V}_{i,j}(t)=\langle i|\hat{V}(t)|j\rangle$ represent the time-dependent field-atom 
interaction between two electronic states $|i\rangle$ and $|j\rangle$, \hbox{$\hat{V}(t)=e\bm p\cdot {\bm{A}}(t)$}  is
the dipole operator in the velocity gauge, ${\bm{A}}(t)=F(t)\sin(\omega t)\,\hat{\bm{e}}_y$ is the vector potential, 
and $F(t)$ is a smooth envelope that peaks at the maximum field strength.
 
Because the laser is tuned near the $2s\to2p$ transition, we can restrict the first transition 
to the $|2p,m=\pm1\rangle$ states. As a result, the amplitudes can be split into two terms
  \begin{eqnarray}
\mathcal{A}^{(3)}_{2s}(\bm p)&=&\mathcal{A}^{(3),+}_{2s}(\bm p)+\mathcal{A}^{(3),-}_{2s}(\bm p),
\label{eq:split}
 \end{eqnarray} 
where $\mathcal{A}^{(3),\pm}_{2s}(\bm p)$ are the three-photon amplitudes with the first transition 
to one of the $|2p,m=\pm1\rangle$ states. These amplitudes take the form
 \begin{eqnarray}
\mathcal{A}^{(3),\pm}_{2s}(\bm p)&=&e^{+i\frac{E_{2s}}{\hbar}t_0}\left(-\frac{i}{\hbar}\right)^2\sum_{j}\int_{t_0}^{\infty}dt_1\int_{t_0}^{t_{1}}dt_{2}e^{i\frac{E-E_{j}}{\hbar}t_1}\nonumber\\
&\times&\hat{V}_{\bm k,j}(t_1) e^{i\frac{E_{j}-E_{2p}}{\hbar}t_2}\hat{V}_{j,2p\pm}(t_2) a^{(1),\pm}_{2s\to 2p}(t_2),
\end{eqnarray} 
where $\hat{V}_{j,2p\pm}(t_2)=\langle j|\hat{V}(t_2)|2p,m=\pm1\rangle$ and $E_{2p}$ is the $2p$ state energy. 
In \eqref{eq:split}, we introduced the one-photon time-dependent transition amplitudes to the $2p$ states
 \begin{eqnarray}
a^{(1),\pm}_{2s\to 2p}(t_2)=-\frac{i}{\hbar}\int_{t_0}^{t_{2}}dt_{3}\hat{V}_{2p\pm,2s}(t_3)e^{i\frac{E_{2p}-E_{2s}}{\hbar}t_3}.
\label{eq:1-phot-amp}
\end{eqnarray} 
For light polarized along the $y$ axis, $a^{(1),+}_{2s\to 2p}=a^{(1),-}_{2s\to 2p}$.

We define the dynamical phase, $\Delta\phi(t)$, by 
 \begin{eqnarray}
a^{(1),\pm}_{2s\to 2p}(t)=|a^{(1),\pm}_{2s\to2p}(t)|e^{i[\phi_0+\Delta\phi(t)]}.
\end{eqnarray} 
Because the detuning, $\Delta\omega=\omega-(E_{2p}-E_{2s})/\hbar$ is very small in our near-resonance 
condition ($\Delta\omega=30$~meV), the dynamical phase, $\Delta\phi(t)\approx (\Delta\omega)t$ varies very slowly. 
For weak fields, when interactions with other states are neglected (two-state model), the dynamical phase 
is identically zero in the resonance limit of $\Delta\omega=0$. 

It is now possible to incorporate $a^{(1),\pm}_{2p\to 2p}(t_2)$ in $\hat{V}(t_2)$ and factor out 
the time-independent phase $\phi_0$. This leads to 
\begin{eqnarray}
\mathcal{A}^{(3),\pm}_{2s}(\bm p)&=&e^{i\beta}\left(-\frac{i}{\hbar}\right)^2\sum_{j}\int_{t_0}^{\infty}dt_1\int_{t_0}^{t_{1}}dt_{2}e^{i\frac{E}{\hbar}t_1}V_{\bm k,j}(t_1)\nonumber\\
&\times&e^{-i\frac{E_{j}}{\hbar}(t_1-t_2)} \bar{V}_{j,2p,\pm}(t_2)e^{-i\frac{E_{2p}}{\hbar}t_2},
\end{eqnarray} 
where $\beta=E_{2s}t_0/\hbar+\phi_0$ and $\bar{V}(t_2)=a^{(1),\pm}_{2s\to 2p}(t_2)\hat{V}(t_2)$. This is the equivalent 
of introducing a new complex field, ${\bar{\bm{A}}}(t)=\bar{F}(t)\exp[{i\Delta\omega t}]\sin\omega t\;\hat{\bm e}_y$, for the second photon transition. 
Therefore, the new field has a frequency shifted by the detuning, $\pm\Delta\omega$, for emission/absorption, respectively, and a new envelope given by 
\begin{eqnarray}
\bar{F}(t)=F(t)|a^{(1),\pm}_{2s\to 2p}(t)|.
\end{eqnarray} 
Because $|\Delta\omega|\ll\omega$ in our near-resonance condition, the frequency shift $\Delta\omega$ 
has a negligible effect on the second transition. In addition, for an adiabatic process as 
in our study, $|a^{(1),\pm}_{2s\to 2p}(t)|\propto F(t)$. Consequently, $\bar{F}(t)\propto F^2(t)$, 
which corresponds to an effective decrease of the pulse intensity and a reduction by two of 
the FWHM for a gaussian envelope. The broadening of the spectral width for a relatively long pulse 
also has a negligible impact on the final phase amplitude.

Therefore, the amplitude is simply rescaled when compared to photoionization directly from the $2p$ 
states with the field ${\bm{A}}(t)$, and we can write
\begin{eqnarray}
\mathcal{A}^{(3),\pm}_{2s}(\bm p)\approx\mathcal{A}_{2s\to2p}\mathcal{A}^{(2),\pm}_{2p}(\bm p),
\end{eqnarray} 
where $\mathcal{A}_{2s\to2p}$ is a momentum-independent complex amplitude.   
The two-photon ionization amplitudes, starting from $|2p,m=\pm1\rangle$, then take the form
\begin{eqnarray}
\mathcal{A}_{2p}^{(2),\pm}&=&\left(-\frac{i}{\hbar}\right)^2\sum_{j}\int_{t_0}^{\infty}dt_1\int_{t_0}^{t_{1}}dt_{2}e^{i\frac{E}{\hbar}t_1}V_{\bm k,j}(t_1)\nonumber\\
&\times&e^{-i\frac{E_{j}}{\hbar}(t_1-t_2)} V_{j,2p,\pm}(t_2)e^{-i\frac{E_{2p}}{\hbar}t_2}. 
\end{eqnarray} 
As a result, we obtain  
\begin{eqnarray}
|\mathcal{A}^{(3)}_{2s}(\bm p)|^2\approx|\mathcal{A}_{2s\to2p}|^2|\mathcal{A}_{2p}^{(2),+}(\bm p)+\mathcal{A}_{2p}^{(2),-}(\bm p)|^2. \nonumber \\
\end{eqnarray} 

In our case, higher-order processes might play a role even in the weak-field regime due to the 
near-resonance condition. However, the above reasoning applies in  exactly the same way for higher-order processes. 
We generally obtain for the $(2n+1)$-photon amplitude
\begin{eqnarray}
\mathcal{A}^{(2n+1),\pm}_{2s}(\bm p)\approx\mathcal{A}_{2s\to2p}\mathcal{A}^{(2n),\pm}_{2p}(\bm p).
\end{eqnarray} 

Finally, summing up the different order terms and introducing $\alpha=|\mathcal{A}_{2s\to2p}|^2$, 
which only depends on the $2s\to 2p$ transition, we obtain Eq.~\eqref{eq:interference}, i.e.
\begin{eqnarray}
|\mathcal{A}_{2s}(\bm p)|^2\approx\alpha|\mathcal{A}^+_{2p}(\bm p)+\mathcal{A}^-_{2p}(\bm p)|^2.
\end{eqnarray} 

Numerically, we found the above relation to be quite robust against increasing the pulse intensity and, in 
particular, at intensities when the population transfer becomes non\-adiabatic. This is explained by 
the fact that \hbox{$\Delta\phi(t)\approx(\Delta\omega)t$} holds at higher intensities if one 
assumes negligible interactions with other states. Surprisingly, as Rabi oscillations become visible 
at slightly higher intensity ($\approx 10^{12}\,$W/cm$^2$), Eq.~\eqref{eq:interference} still remains 
relatively accurate. This is likely due to the fact that higher-order terms are able to describe 
population transfer between the $2s$ and $2p$ states. As Rabi 
oscillations take place, the effective envelope $\bar{F}(t)$ oscillates at the Rabi frequency~$\Omega$, 
such that the pulse obtains two additional frequency components $\omega\pm\Omega$. This represents an 
alternative interpretation of the Autler-Towns splitting besides the dressed-state picture.

When the intensity keeps increasing, the perturbative expansion will break down and other 
states, besides $2p$, will start contributing to the first transition step.  As a result, the 
approximation \eqref{eq:split} will ultimately become inaccurate.

\end{document}